\documentclass[onecolumn, floatfix, showpacs, superscriptaddress, aps, 10pt, prb]{revtex4-2}
\usepackage{amsmath}
\usepackage{graphicx}
\usepackage{dcolumn}
\newcolumntype{d}[1]{D{.}{.}{#1}}
\usepackage{bm}
\usepackage{epstopdf}
\usepackage{graphicx}
\usepackage{stmaryrd}
\usepackage{tikz}
\usepackage{lipsum}
\usepackage{pgf}
\usepackage{float}
\usepackage{hyperref}
\usepackage{tabularx}
\usepackage{appendix}
\usepackage{subfigure}
\usepackage[section]{placeins}
\usepackage[none]{hyphenat} 
\newcommand{\be}{\begin{equation}} 
\newcommand{\ee}{\end{equation}} 
\newcommand{\bea}{\begin{eqnarray}} 
\newcommand{\eea}{\end{eqnarray}} 
\newcommand{\bqa}{\begin{eqnarray}}
\newcommand{\eqa}{\end{eqnarray}}
\newcommand{\mb}{\mathbf}
\newcommand{\mc}{\mathcal}
\newcommand{\nn}{\nonumber \\}

\newcommand{\w}{\omega}

\newcommand{\bwt}{\begin{widetext}}    
\newcommand{\ewt}{\end{widetext}}     

\newcommand{\figtato}
{\begin{figure}[htbp]
        \centering
        \includegraphics[width=5cm]{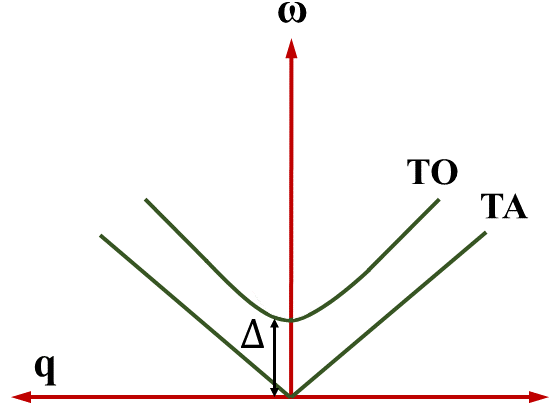}
           \caption{A schematic description of proximity of the transverse acoustic (TA) and transverse optical (TO) phonon modes associated with (incipient) ferroelectric transition in a typical quantum paraelectric. Here $\Delta = \Delta_0 + aT^2$ is the temperature dependent gap between the TA and TO phonons, where $T$ is the temperature, $a$ is material dependent parameter, and $\Delta_0$ is the phonon energy gap at $T=0$.}
           \label{fig:figtato}
	\end{figure}
}
\newcommand{\figdiaen}
{\begin{figure}[htbp]
        \centering
        \includegraphics[height=2cm,width=10cm]{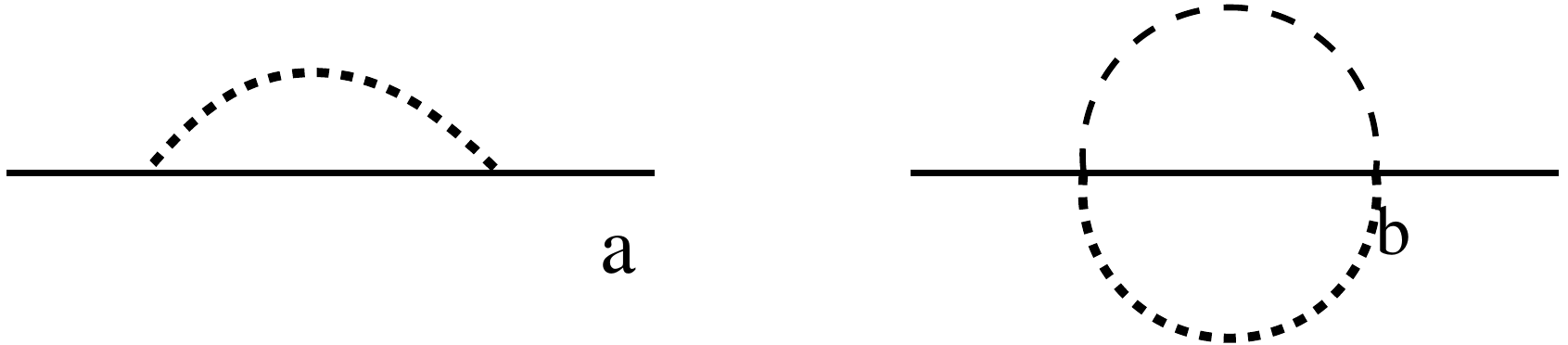}
           \caption{a): Lowest order three phonon process due to the cubic anharmonic term where the quantum fluctuations of the relevant optic branch contribute the most and b) Second order four phonon process due to the quartic anharmonic term. Here the dotted and the dashed curves represent two different kind of phonons.}
           \label{fig:figdia-en}
	\end{figure}
}
\newcommand{\figdiamass}
{\begin{figure}[htbp]
        \centering
        \includegraphics[height=2cm,width=6cm]{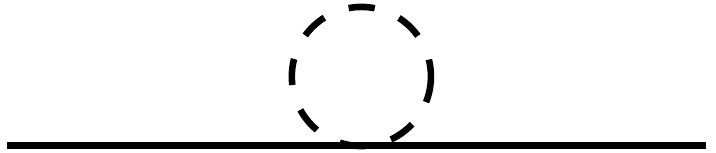}
           \caption{Mass enhancement due to the lowest order quartic anharmonic effect which leads to Eq.~\ref{eq:dispersion} with renormalized gap $\Delta_0\sim T^2.$}
           \label{fig:figdia-mass}
	\end{figure}
}

\newcommand{\figmain}
{\begin{figure}[htbp]
        \centering
        \includegraphics[height=5cm,width=7cm]{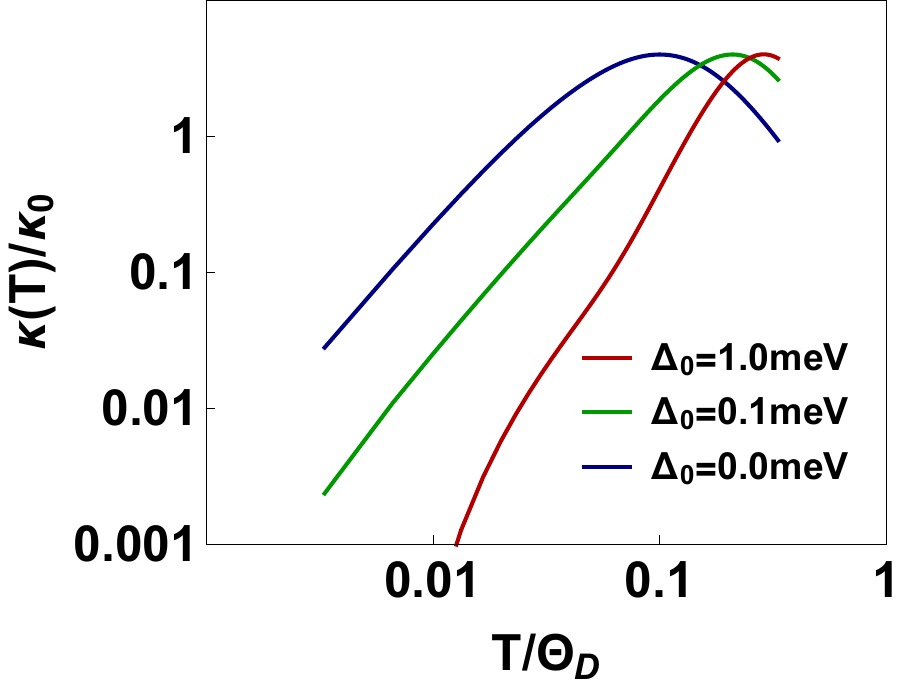}
           \caption{Power law behavior of the optical phonon contribution to the thermal conductivity near the quantum critical point as obtained in the present formalism.}
           \label{fig:figmain}
	\end{figure}
}

\begin{document} 
\baselineskip 12pt 
\title {Optical phonon contribution to the thermal conductivity of a quantum paraelectric }
\author{Pankaj Bhalla}
\affiliation{Beijing Computational Science Research Center, Beijing, 100193, China}
\author{Nabyendu Das}
\affiliation{Department of Physics, The LNM-Institute of Information Technology, Jaipur 302031, India
} 
\date{\today}

\begin{abstract}
Motivated by recent experimental findings, we study the contribution of a quantum critical optical phonon branch to the thermal conductivity of a paraelectric system.  We consider the proximity of the optical phonon branch to transverse acoustic phonon branch and calculate its contribution to the thermal conductivity within the Kubo formalism. We find a low temperature power law dependence of the thermal conductivity as $T^{\alpha}$, with $1 < \alpha < 2$, (lower than $T^3$ behavior) due to optical phonons near the quantum critical point. This result is in accord with the experimental findings and indicates the importance of quantum fluctuations in the thermal conduction in these materials.
\end{abstract}
\maketitle
\section{Introduction} 
In the last few decades, the study of quantum paraelectrics such as SrTiO$_3$ and KTaO$_3$ has become a topic of considerable interest \cite{das_JPCM2009, rowley_NP2014, palova_PRB2009, das_PLA2012, das_IJMPB2013, das_MPLB2014}. In general, interest in the quantum fluctuation induced novel states is on rise as it can potentially lead to technological advancements. The main feature of these materials is a suppressed long range ferroelectric order down to the zero temperature due to quantum fluctuations. It is found that the dielectric constant saturates at a very high value on lowering the temperature. Thus these materials in their pristine state may be considered to be on the verge of a phase transition near $T=0$. In principle, the ferroelectric transition can be induced by some non thermal parameters such as pressure, chemical composition, etc. at a zero temperature. A continuous zero temperature transition without any discontinuity in the order-parameter variations are termed as quantum critical point. The ``zero temperature transition" or a ``quantum critical point" are not just theoretical concepts with academic interests only. Once the non-thermal parameters in a system are tuned to value that lead to a zero temperature continuous transition, they determine the finite temperature properties of a system in terms of quantum critical fluctuations of the associated degrees of freedom. In such cases, the finite temperature properties shows some universal behaviour which are independent of many details of the microscopic interations \cite{sachdev_book}.  A set of dielectric materials such as SrTiO$_3$, KTaO$_3$  associated with such phenomena as a result of quantum critical polarization fluctuations are coined as quantum paraelectrics. In these materials (suppressed) ferroelectric transition is associated with a nearly soft zone center optical mode and in the low temperature regime, many of the thermodynamic and transport properties can be described by quantum fluctuations of the associated optic branch and its interactions with other degrees of freedom. Such novel dielectric behavior was first reported by M\"uller et al. \cite{muller_PRB1979} in strontium titanate compounds. In recent times, pressure induced quantum phase transitions and associated quantum critical behavior have been experimentally confirmed \cite{rowley_NP2014}. A lot of theoretical and experimental activities in this direction to describe the dielectric behavior of these systems have been successfully taken place in the recent times \cite{palova_PRB2009, conduit_PRB2010, narayan_NM2019, li_sci2019, ang_PRB2000, khaetskii_JPCM2020, geirhos_PRB2020, xing_PRL2020, kustov_PRL2020, nataf_NRP2020, sim_PRL2021}.\\ 
To explore consequences of quantum fluctuations in these materials, the study of heat transport properties is also equally important like their dielectric behavior. This may shed further light on the behavior of quantum paraelectrics at or near the quantum critical point. In an insulator heat conduction is mainly driven by the flow of phonons. Optical phonons have higher energies $\sim$ Debye temperature and they often lead to novel thermal conductivity behavior at the said temperature regime \cite{zhen_Carbon2015, zou_APL2019}. However, at low temperature thermal conductivity is dominated by the flow of the acoustic phonons. In case of a quantum paraelectric, zone center transverse optic mode is nearly soft and the phonons in the respective transverse optical (TO) branch have lowered energy compared to the usual insulators. This is to mention that LO phonons in this materials still have higher energy owing to the long range dipolar interaction \cite{coak_PRB2019}.This scenario is observed in experiments \cite{shirane_PR1967, farhi_EPJB2000} and schematically shown in Fig.~\ref{fig:figtato}. Thus it is plausible that optical phonons in a quantum paraelectric shape the low temperature behavior in certain systems, such as KTaO$_3$. 
\figtato \\
Being high electronic energy gap semiconductors, electronic exciations are gapped out at a low temperature and transport properties in these materials are governed by the interactions between lowered energy optical phonons with the acoustic phonons \cite{martelli_PRL2018, yang_PRL2020, li_PRL2020} and other degrees of freedom like impurities, structural domains\cite{kustov_PRL2020}, etc.\\
The basic understanding of the phonon thermal conductivity is as follows. In a classical kinetic theory estimate, the thermal conductivity $\kappa\sim v_{\text{ph}}c_V\lambda_{\text{ph}}$, where $v_{\text{ph}}$, $c_V$, and $\lambda_{\text{ph}}$ are the phonon velocity, phonon specific heat and the phonon mean free path respectively \cite{ziman_book, tritt_book, berman_book}. Thermal conductivity being a transport coefficient, depends on the relaxation of the phonons due to its interactions with other degrees of freedom and or anharmonicities. The same is encoded in the $\lambda_{\text{ph}}=v_{\text{ph}}\tau_{ph}$, $\tau_{ph}$ is the relaxation time. At low temperature and for acoustic phonon dominated scenario, $c_V\sim T^3$ and $\lambda_{\text{ph}}$ is determined by the scattering with the impurities and is temperature independent. Thus $\kappa\sim T^3$. Experimental findings on this materials also support such $T^3$ behavior of the specific heat \cite{bourgeal_Ferr1988}. Thus any departure from the said $T^3$ behavior of the thermal conductivity, especially in a regime where some other interacting degrees of freedom become active, indicates the presence of some new scattering mechanism which results in frequancy and temperature dependence of the mean free path or the scattering rate. \\
A recent experiment shows a few such novel behavior of the thermal conductivity in quantum paraelectrics. For SrTiO$_3$ there is a stronger than $T^3$ increase of the thermal conductivity at low temperature. Poiseuille flow of the phonons has been argued as a possible explanation of this behavior. On the other hand it has been reported that in case of KTaO$_3$, there is a weaker than $T^3$ behavior for the thermal conductivity \cite{martelli_PRL2018}. The later behavior is typically attributed to the mutual interactions between more than one low lying modes \cite{gurzhi_PU1968}. These experimental results suggest that quantum paraelectrics are wonderful playgrounds to explore them. We put forward such an explanation in terms of the thermal conduction of the optical phonons and its interaction with the acoustic phonons which are the relevant degrees of freedom in this system and in the temperature regime of our interest.\\  
Our main result can be summarized as follows. Within a minimal model with coupled TA-TO phonons, it is shown that near the quantum critical point, the temperature variation of the thermal conductivity due the nearly soft optical phonon follows a power law $T^\alpha$, $2>\alpha>1$ and its numerical value can increase an order of magnitude. The same is shown in Fig.~\ref{fig:figmain}. 
\figmain \\
This paper is organized as follows. First, we provide the mathematical description of the anharmonic effects in Sec.~\ref{sec:phonondecay} by introducing the model Hamiltonian terms of the order of cubic and quartic in phonon displacement functions. Then, we derive the expressions of the decay constants and thermal conductivity in Sec.~\ref{sec:thermalconductivity}. In Sec.~\ref{sec:results}, we present the numerical results. In Sec.~\ref{sec:discussion}, we discuss our results and conclude.
\section{Phonon decay due to anharmonicity} 
\label{sec:phonondecay}
In a perfect crystal, harmonic phonons do not decay due to their infinitely long lived nature and lead to infinite thermal conductivity \cite{mermin_book}. We need to introduce symmetry permitted anharmonic interactions to study the finite lifetime of the phonons and hence the associated transport porperties \cite{klemens_PR1966, chan_PRB1962} In the present work, we also focus on the study of anharmonic effects on the thermal conductivity of quantum paraelectrics.\\ 
For a system in which atoms are displaced from their equilibrium positions, the total Hamiltonian can be written as 
\bea
H &=& H_{\text{har}} + H_{\text{anhar}},
\eea
where $H_{\text{har}}$ corresponds to a free phonon part of the Hamiltonian and $H_{\text{anhar}}$ represents the corrections to the Harmonic approximation. The Harmonic part in terms of the phonon coordinates can be expressed like
\bea
H_{\text{har}} &=&\sum_{\alpha\mb{q}} \frac{1}{2}\w_{\mb{\alpha q}}A_{\alpha \mb{q}}A_{\alpha\mb{-q}}.
\label{eq:H0}
\eea
Here $A_{\alpha\mb{q}} = a^\dag_{\alpha\mb{q}}+a_{\alpha \mb{q}}$ is the sum of the creation $(a^\dag_{\alpha\mb{q}})$ and the annihilation $(a_{\alpha\mb{q}})$ phonon operators, $\omega_{\alpha\mb{q}}$ is the phonon dispersion for the $\alpha$-th branch. The latter quantity for small wavelengths can have form 
\be
\w_{\alpha q}= \sqrt{{\Delta_{\alpha 0}}^2+c_\alpha^2q^2},
\label{eq:dispersion}
\ee
having $\Delta_{\alpha 0}$ the phonon energy gap, $c_\alpha$ a sound velocity and $q$ a phonon wave vector. 
The dispersion correspond to the phonon branch can be associated with the gap. For the case $\Delta_0=0$, the dispersion corresponds to acoustic phonon and for  $\Delta_0\neq 0$ it represents optical phonon. Since we are considering a minimal model with one TA and one TO branch, we drop the branch index $\alpha$ for further discussions and $\Delta_0$ represents the energy gap for the zone center TO phonon at $T=0$.
Further, the ferroelectric instability in materials having anharmonic effects is determined with $q=0$ optical phonon or $\w_{q} = \Delta_0$. The ferroelectric quantum critical point, it is determined by $\Delta_0=0$ at $T=0$ for the optic branch. At this quantum critical point, the phonon energy gap for the optical branch shows quadratic temperature dependent phonon dispersion at finite temperature such as $\Delta = \Delta_0 + a T^2$ having $a$ as a material dependent parameter  \cite{rowley_NP2014, das_JPCM2009} as depicted in Fig.~\ref{fig:figtato}.
\figdiaen \\
In the anharmonic part $H_{\text{anhar}}$ of the total Hamiltonian, we consider two lowest order terms of the following  forms.
\bea
H_3&=&\sum_{\alpha \beta\gamma\mb{k}_1\mb{k}_2\mb{k}_3 } V_{\alpha \beta \gamma } (\mb{k}_1, \mb{k}_2, \mb{k}_3)  A_{\alpha ,\mb{k}_1}A_{\beta ,\mb{k}_2}A_{\beta ,\mb{k}_3}\nn
H_4&=&\sum_{\alpha \beta\gamma\mb{k}_1\mb{k}_2\mb{k}_3 } V_{\alpha \beta \gamma \delta} (\mb{k}_1, \mb{k}_2, \mb{k}_3, \mb{k}_4) A_{\alpha ,\mb{k}_1}A_{\beta ,\mb{k}_2}A_{\beta ,\mb{k}_3}A_{\delta ,\mb{k}_4} \nn 
\eea
Here $V$ in both terms represents the anharmonic coefficients and the relative weights of these coefficients depend on the symmetry of system. The Hamiltonian $H_{3}$ corresponds to the three phonon process and $H_{4}$ to the four phonon process. The lowest order processes form both types of anharmonicities which involve momentum transfer between two kind of phonons are shown in Fig.~\ref{fig:figdia-en}. Here the first order effect from the fourth order anharmonicity does not associated with any net momentum transfer. It contributes to the quadratic temperature dependence to the phonon energy gap as shown in Fig.~\ref{fig:figdia-mass}.  
\figdiamass
\subsection{Three Phonon process}
For the three phonon process, decay rate or the inverse lifetime of a phonon at the phonon frequency due to the cubic anharmonicity and at a temperature $T$ is given as:
\be
\Gamma^{3}(\w , T)=18\pi\sum_{\mb{k}_1\mb{k}_2\lambda_1 \lambda_2}\vert V_{\alpha \beta}(-\mb{k}\lambda;\mb{k}_1\lambda_1;\mb{k}_2\lambda_2)\vert^2 \bigg\{ (n_{\w_1}-n_{-\w_2})\delta_{12}^{s}
+(n_{\w_1}-n_{\w_2})\delta_{12}^{d} \bigg\}.
\ee 
Here $\w_i\equiv \left\lbrace \w(\mb{k}_i\lambda_i)\right\rbrace$, $n_{\w_i}=(e^{\beta\w_i}-1)^{-1}$ is the Bose-Einstein distribution function having $\beta$ a inverse of the temperature, and the dirac delta functions denote $\delta_{12}^{s} = \delta_{\w-\w_1-\w_2}-\delta_{\w+\w_1+\w_2}$ and $\delta_{12}^{d}=\delta_{\w+\w_1-\w_2}-\delta_{\w-\w_1+\w_2}$. Considering the anharmonic coefficient $\vert V(-\mb{k}\lambda;\mb{k}_1\lambda_1;\mb{k}_2\lambda_2)\vert^2=\frac{V_0^2}{\w(-\mb{k}\lambda)\w(\mb{k}_1\lambda_1)\w(\mb{k}_2\lambda_2)}$\cite{peierls_book} and then converting the summations into integrals, we have
\be
\Gamma^{3}(\w , T)=18\pi \frac{V_0^2}{\w}\int d\w_1 \rho_{1}^{\w_1} \bigg\{ \rho_{2,\w}^{\w_1}(n_{\w_1}-n_{-\w +\w_1}) +\rho_{2,\w}^{-\w_1}(n_{\w_1}-n_{\w+\w_1})- \text{terms with} (\w \rightarrow -\w )\bigg\},
\ee 
where $\rho_{i, \w}^{\pm \w_1} = \rho_{i}(\w \mp \w_1)/(\w \mp \w_1)$ having $\rho_{i}(\w)$ the density of states. For a certain phonon branch, it takes a form $\rho_i({\w})\sim$ $\w\sqrt{\w^2-\Delta^2}$ for $\w>\Delta$ and zero in other cases. Here we consider $\Delta=\Delta_0+ a T^2$ with $a$ eV$^{-1})$ as a constant for optic phonon which is not explicitly calculated here and taken from other references \cite{das_JPCM2009, rowley_NP2014} and shown in Fig.~\ref{fig:figtato}.\\
To simplify the expression of decay rate, we have used $\rho_{1}(\omega)$ for optic phonon and $\rho_{2}(\omega)$ for acoustic phonon. This yields $\Gamma^{3}(\w ,T)$
\be
\Gamma^{3}(\omega, T) =\frac{18\pi }{4\pi^4} \frac{V_{0}^2}{\omega} \int d\w_1 \sqrt{\w_1^2 - \Delta^2}\bigg\{ (\w-\w_1)\big[n_{\w_1}-n_{-\w+\w_1}\big] + (\w+\w_1)\big[n_{\w_1}-n_{\w+\w_1}\big] \bigg\}.
\label{eqn:DR3}
\ee
This is an expression for the decay rate of the optical phonon due to three phonon process.

\subsection{Four Phonon process}
Due to the fourth order anharmonicity, the phonon decay rate can be expressed as
\bea \nn
\hspace{-2.5cm}\Gamma^{4}(\omega,T) =& 96\pi \sum_{\mb{k}_1\lambda_1} \sum_{\mb{k}_2\lambda_2}\sum_{\mb{k}_3\lambda_3} \vert V(-\mb{k}\lambda;\mb{k}_1\lambda_1;\mb{k}_2\lambda_2;\mb{k}_3\lambda_3)\vert^2  \nn
& \quad \bigg\{ (-n_{-\w_{1}}n_{-\w_{2}}n_{-\w_{3}} - n_{\w_{1}}n_{\w_{2}}n_{\w_{3}}) \delta_{123}^{s}+ 3( n_{\w_{1}}n_{-\w_{2}}n_{-\w_{3}} + n_{-\w_{1}}n_{\w_{2}}n_{\w_{3}}) \delta_{123}^{d}\bigg\} \nn ,
\eea
where $\delta_{123}$ represents the dirac delta function due to the combination of different phonon frequencies. The corresponding factors with superscript $s$ and $d$ are $\delta_{\w -\w_{1} - \w_{2} -\w_{3}} -  \delta_{\w + \w_{1} + \w_{2} +\w_{3}}$, and $\delta_{\w + \w_{1} - \w_{2} -\w_{3}} - \delta_{\w - \w_{1} + \w_{2} + \w_{3}}$ respectively.\\
Assuming the anharmonic coefficient for this process in a similar fashion to the three phonon process $\vert V(-\mb{k}\lambda;\mb{k}_1\lambda_1;\mb{k}_2\lambda_2; \mb{k}_3\lambda_3)\vert^2=\frac{V_1^2}{\w(-\mb{k}\lambda)\w(\mb{k}_1\lambda_1)\w(\mb{k}_2\lambda_2)\w(\mb{k}_3\lambda_3)}$ and simplifying the above equation, the the decay rate for four phonon process can been written as
\bea 
\hspace{-2.5cm}\Gamma^{4}(\omega,T) =& 96\pi \frac{V_{1}^{2}}{\w} \int d\w_1 \rho_{1}^{\w_1} \int d\w_2 \rho_{2}^{\w_2} \bigg\{ \rho_{2}^{-\w_{1}-\w_{2}} (-n_{-\w_1}n_{-\w_2}n_{-\w + \w_1 +\w_2} - n_{\w_1}n_{\w_2}n_{\w - \w_1 -\w_2}) \nn [2ex]
& +3 \rho_{2}^{\w_1 -\w_2} (n_{\w_1}n_{-\w_2}n_{-\w - \w_1 +\w_2} + n_{-\w_1}n_{\w_2}n_{\w + \w_1 -\w_2}) - \text{terms with} (\w \rightarrow -\w) \bigg\}\nn
\label{eqn:DR4}
\eea
Here we consider restrict us to the simplest model which captures the essential aspects of the quantum fluctuations associated with the nearly soft optic mode and consider $\rho_{1}(\w)$ for optic phonon and other density of states for the case of acoustic phonon branches.
\section{Thermal conductivity}
\label{sec:thermalconductivity}
In order to compute the thermal conductivity, we start with its expression based on the Kubo approach \cite{mahan_book} and is given by
\bea
\kappa(\w,T)&=& \frac{\kappa_0}{T}\sum_{\mb{k} } \w^2(\mb{k})
v^2_{\mb{k}}\int_{-\infty}^{\infty}d\w' f_T(\w, \w') \mc{S}(\w,\w').\nn
\label{eqn:TC}
\eea
Here $f_T(\w, \w')=\frac{n_{\w'}-n_{\w+\w'}}{\w}$ is the thermal weight and the spectral weight function $\mc{S}(\w,\w')$ is defined as $\mc{S}(\w,\w')=\mc{A}(\mb{k},\w')\mc{A}(\mb{k},\w+\w')$. Further in zero frequency limit, the above equation can be written as
\be 
\kappa(\w= 0, T)=\frac{\kappa_0}{T}\sum_{\mb{k} } \w^2(\mb{k})
 v^2_{\mb{k}}\int_{-\infty}^{\infty}d\w'\frac{\partial n(\w')}{\partial \w'}\mc{A}^2(\mb{k},\w').
\label{eqn:TCZ} 
\ee
Here the spectral function $\mc{A}(\mb{k}, \w)$ is defined as the imaginary part of the phonon propagator and multiplied by a factor $1/\pi$ i.e.,
\bea 
\mc{A}(\mb{k},\w)&=&-\frac{1}{\pi}  Im \mc{D}(\w,\mb{k})\nn
&=& -\frac{1}{\pi}\frac{2\w(\mb{k})\Gamma(\mb{k}, \w)}{\left(\w^2-\w^2(\mb{k})\right)^2+4\w(\mb{k})^2\Gamma(\mb{k}, \w)^2}.
\eea 
Using the expressions for decay rate Eqs.~(\ref{eqn:DR3}) and (\ref{eqn:DR4}) and the phonon dispersion in the above equation, the thermal conductivity Eq.~(\ref{eqn:TCZ})can be calculated. This calculation is within linear response theory by beyond relaxation time approximation. We calculate these expressions numerically and discuss the behavior of relevant quantities in the next section.
\begin{figure}[t]
	\centering
	\subfigure[noonleline][]
	{\label{fig:scattdccomp}\includegraphics[height=5cm,width=7cm]{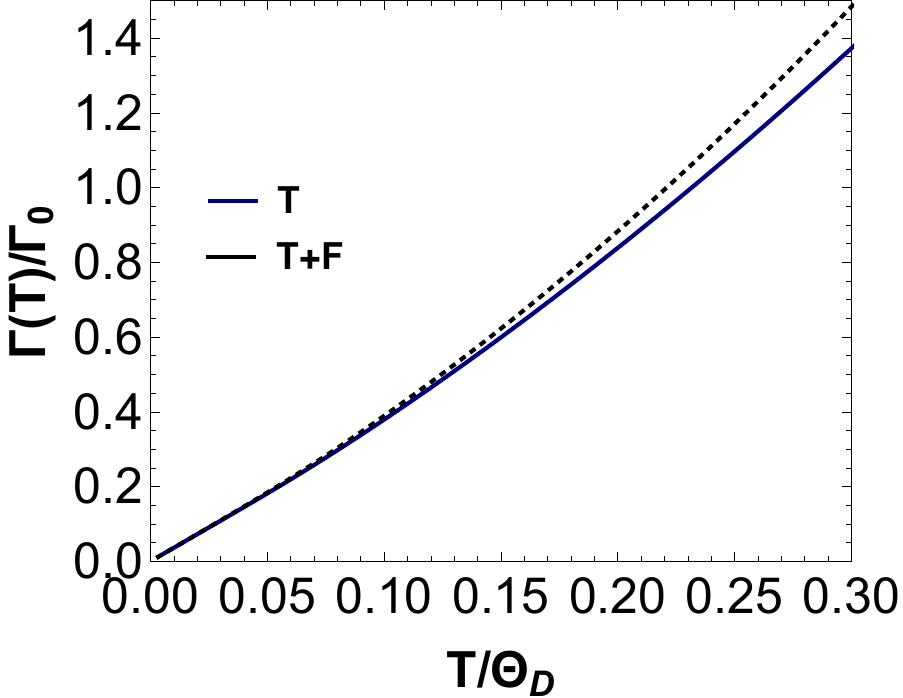}}
	\subfigure[noonleline][]
	{\label{fig:scattaccomp}\includegraphics[height=5cm,width=7cm]{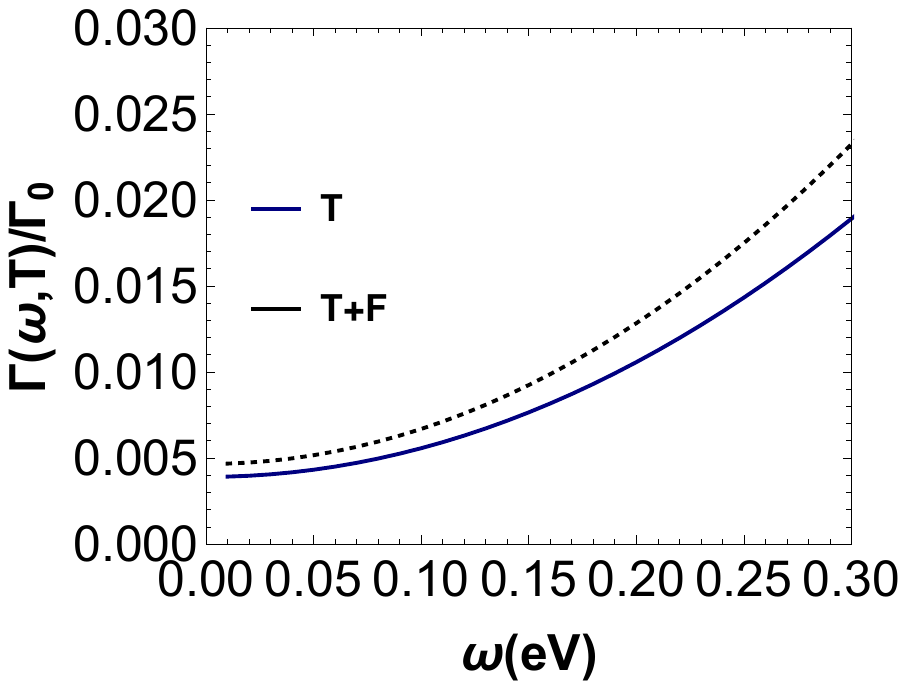}}
	\caption{(a). Normalized temperature dependent thermal decay rate as a function of $T/\Theta_{D}$ having $\Theta_D$ a Debye's temperature corresponds to three (T) and four phonon processes. Here the solid blue curve refers to the three phonon contribution and dotted black curve to the three and four  phonon (T+F) combination.(b). Decay rate at a fixed temperature as a function of frequency.}
	\label{figure1}
\end{figure}
\section{Results}
\label{sec:results}
In Fig.~\ref{fig:scattdccomp}, we plot the normalized thermal decay rate $\Gamma(T)/\Gamma_0$ using Eqs.~\ref{eqn:DR3} and \ref{eqn:DR4} as a function of $T/\Theta_D$ having $\Theta_D$ a Debye's temperature at a gap $\Delta_0=0$. Here, we observe that the scattering rate increases with temperature. Both three and phonon processes contribute to rise in $\Gamma(T)/\Gamma_0$. However, the contribution from four phonon process is quite small as shown by black dotted curve in Fig.~\ref{fig:scattaccomp}. Further it shows nonlinear behavior at temperature greater than $0.1$ Debye temperature. In frequency dependent case, it has been observed that the decay rate varies nonlinearly with the freqeuncy, shown in Fig.~\ref{fig:scattaccomp}.
\begin{figure}[t]
	\centering
	\subfigure[noonleline][]
	{\label{fig:scattdc}\includegraphics[height=6cm,width=7cm]{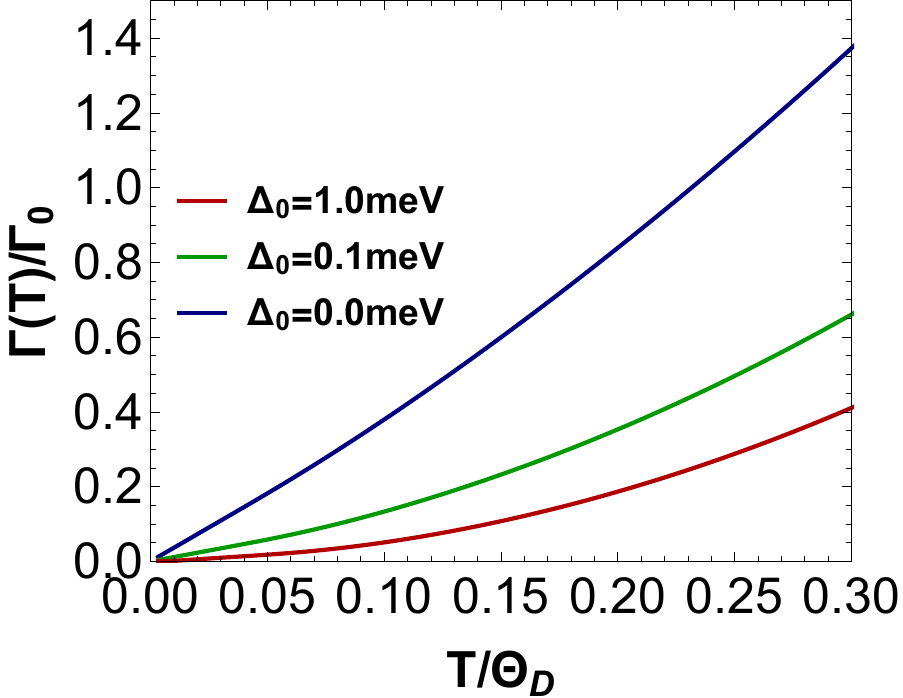}}
	\subfigure[noonleline][]
	{\label{fig:scattac}\includegraphics[height=6cm,width=7cm]{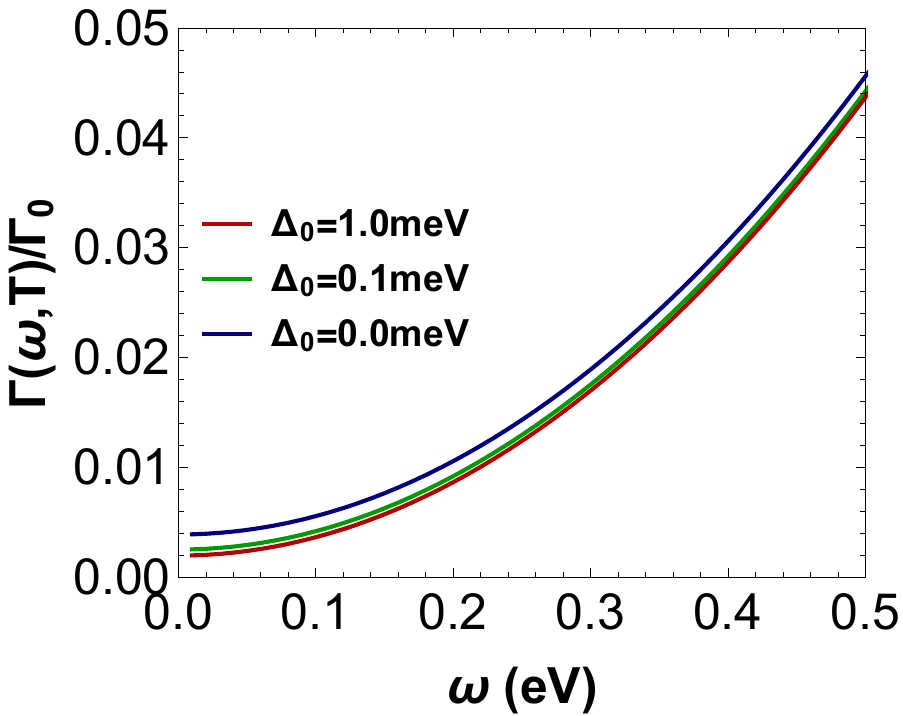}}
	\caption{Plot for the normalized thermal decay rate of the optical phonon as a function of (a): scaled temperature $T/\Theta_D$, (b): frequency  at different gap values $\Delta_0$ such as $0.0$, $0.1$, and $1.0$ meV.}
	\label{figure1}
\end{figure} 
\begin{figure}[htbp]
\centering
\includegraphics[width=7cm,height=5cm]{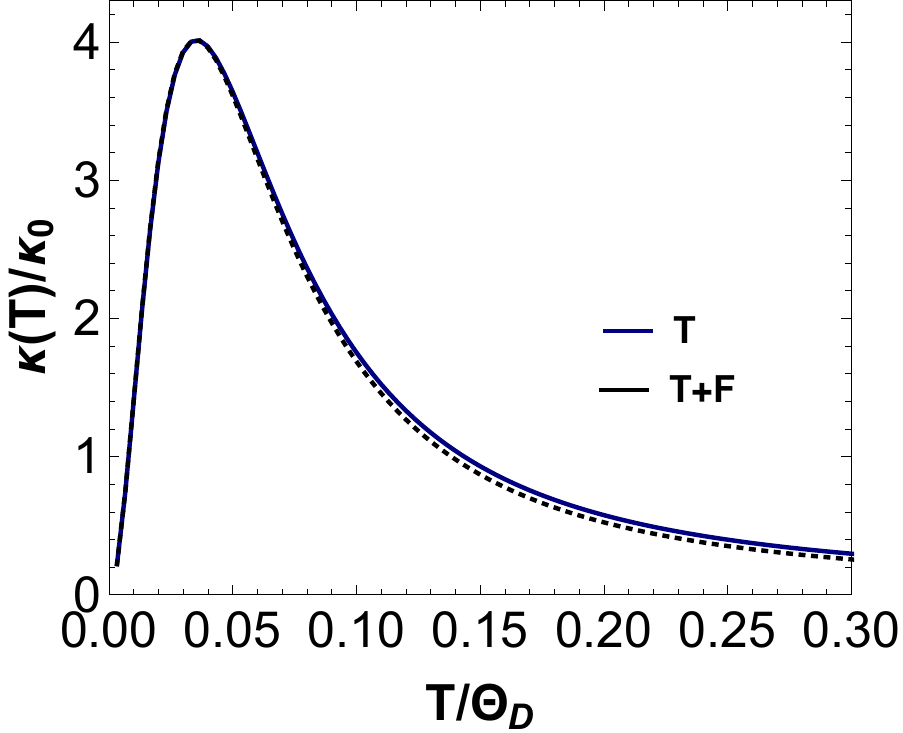}
\caption{Thermal conductivity of a paraelectric due to three (T) phonon process and the combination of three and four (T+F) processes. The curve has been plotted at the quantum critical point $\Delta_0=0$. }
	\label{fig:thermaldccomp}
\end{figure} \\
Refering Fig.~\ref{fig:scattdc} for the temperature dependent normalized scattering rate at different gap values, we have following findings. Firstly, the increase in the gap $\Delta_0$ suppresses $\Gamma(T)/\Gamma_0$. Physically, this arises due to the less number of excitations of phonon from both acoustic and optical branches. Secondly, on moving away from quantum critical point $\Delta_0=0$, the contribution made by acoustic phonons decreases which results the less decay rate. In the frequency dependent case Fig.~\ref{fig:scattac}, the variation of scattering rate at different gap is not clearly visible due to the less magnitude of gap as compared to frequency scale.\\ 
The thermal conductivity of a paraelectric material is presented in Fig.~\ref{fig:thermaldccomp} for different phonon processes. At low temperatures, four phonon process does not show appreciable contribution to the thermal conductivity using fitting parameters $a=0.01$ eV$^{-1}$ and $V_1$ and $V_0 = 1000$eV$^{-1}$. Our choices of parameter is such that the maximum value of the optical phonon decay rates which $\sim \Gamma_0\sim $ a few meV. This choice is in accord with experimental findings on the spectral width of the optical phonons due to TA-TO interactions \cite{axe_PRB1970, shirane_PR1967} and is well justified for the temperature regime of our interest. Even with same or slightly higher values of $V_1$, the fourth order anharmonicity  gives a little contribution which indicates that third order ahnarmonicity effects play the most significant role in the thermal conduction as expected.

This is to mention that usually four phonon process is considered to be important near the Debye temperature \cite{glassbrenner_PR1964}.  However we had an anticipation that owing to the presence of a nearly soft optic phonon which has significant amount of excitation even at low temperature, it might turnout to be important.  Our calculation rules out that possibility.

Here the plot of $\kappa(T)/\kappa_0$ vs $T/\Theta_D$ shows a peak around a temperature 
\begin{figure}[htbp]
\centering
\includegraphics[width=7cm,height=5cm]{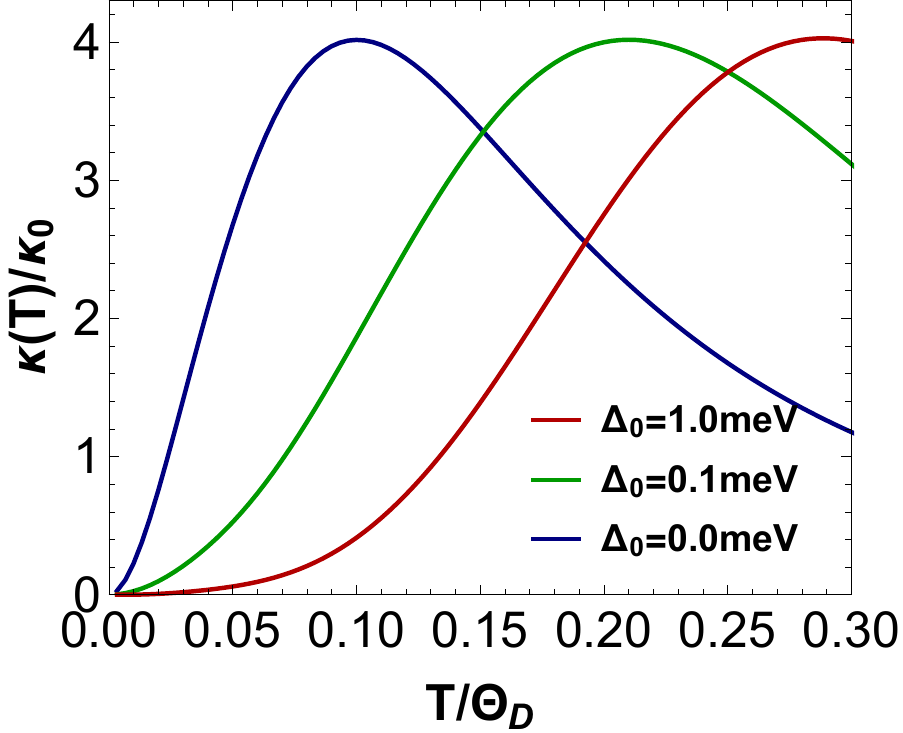}
\caption{Plot for normalized thermal conductivity with $T/\Theta_D$ at different values of gap. The blue curve corresponds to the quantum critical scenario.}
\label{fig:thermal}
\end{figure}
$T \approx 0.075\Theta_D$ and then decays towards the zero as the temperature is lowered. Further, the broadening of the curve is decided by the decay rate $\Gamma(T)$ whose magnitude depends on the strength of the anharmonic coefficients. To look at its behavior away from a quantum critical point, we plot thermal conductivity at different gap values in Fig.~\ref{fig:thermal} as a function of $T/\Theta_D$. We see that at points away from the quantum critical point, the optical phonon does not conduct, resulting vanishing contribution to the thermal conductivity at low temperatures. Also the peak shifts towards the high temperatures as shown in Fig.~\ref{fig:thermal} corresponding to $\Delta_0 = 0.1$, $1$ meV. These results are qualitatively in agreement with experimental results \cite{martelli_PRL2018} as discussed.
\section{Discussion and Conclusion}
\label{sec:discussion}
Our results can be understood as follows. Like acoustic phonons, near quantum critical point optical phonon density is also strongly temperature dependent. Thus the decay rate of the optical phonon is strongly temperature dependent as plotted in Fig.~\ref{fig:scattdc}. We may describe the temperature dependence as $\sim T^{1+x}$ where $0<x<1$. A finite temperature scaling description should give $c_v\sim T^3$ for a quantum critical optical phonon which is same as the acoustic phonons. Considering the above and following the classical expression, at low temperature, $\kappa\sim$ $c_V/\Gamma\sim$ $T^{3-1-x}$ $\sim T^{\alpha}$ $\,\, 2>\alpha=(2-x)>1$. The same has been obtained in our study and shown in the log plot for thermal conductivity in Fig.~\ref{fig:figmain}. Though we do not have a closed analytic expression for the thermal conductivity, the power law obtained here is universal in a sense that is determined by only available energy scale $k_BT$, spatial dimension, symmetry of the order parameter and the dynamical critical exponent $z=1$ and do not depend on the specific values of the interaction parameters. Similar behavior has been obtained in various model calculations on different quantum critical systems: e.g. the loop contribution in 2D superconductor-diffusive metal with z=2, $\kappa\sim \ln T$ \cite{podolsky_PRB2007}.\\
Other important finding form the numerical evaluation of the expression for thermal conductivity is near the critical point, thermal conductivity due to optic phonons enhanced by an order of magnitude than when it is away from a quantum critical point. Thus we propose that TA-Quantum critical TO scattering introduces an additional channel for thermal conduction which may dominate over other contributions in a certain temperature regime. As the related experimental results suggests, the results seem to be relevant for KTaO$_3$ \cite{martelli_PRL2018}, where any structural transition is absent. Our analysis paves a way for addressing quantum critical aspects of thermal conductivity and thermal hall effects in quantum paraelectrics \cite{li_PRL2020, chen_PRL2020} and may be extended to bicritical scenarios in a certain multiferroic sytems \cite{das_PLA2012, morice_PRB2017, narayan_NM2019}.   

\section*{Acknowledgments}
PB acknowledges the National Key Research and Development Program of China (grant No. 2017YFA0303400), China postdoctoral science foundation (grant no. 2019M650461) and NSFC grant no. U1930402 for financial support.

%

\end{document}